\documentstyle[epsf, rotate]{elsart}
\begin{document}
%
%
%
%
 
\begin{frontmatter}
  \title{Monte Carlo Simulation of Magnetic Systems in the Tsallis Statistics}
  \author{A.  R.  Lima\thanksref{email}}
  \thanks[email]{e-mail:arlima@if.uff.br} \address{Instituto de F\'{\i}sica,
    Universidade Federal Fluminense Av.  Litor\^anea, s/n$^o$ - 24210-340
    Niter\'oi, RJ, Brazil} \author{J. S. S\'a Martins\thanksref{email3}}
  \thanks[email3]{e-mail: jorge@cires.colorado.edu} \address{Colorado
Center
    for Chaos and Complexity, University of Colorado, 80309 Boulder, CO, USA}
  \author{T.  J.  P. Penna\thanksref{email2}} \thanks[email2]{e-mail:
    tjpp@if.uff.br} \address{Instituto de F\'{\i}sica, Universidade Federal
    Fluminense Av.  Litor\^anea, s/n$^o$ - 24210-340 Niter\'oi, RJ, Brazil}
  \address{Center for Polymer Studies, Boston University, 02215 Boston, MA,
    USA}

\begin{abstract}

We apply the Broad Histogram Method to an  Ising system in the context
of  the recently  reformulated Generalized  Thermostatistics,  and  we
claim it to be a very efficient simulation tool for this non-extensive
statistics. Results are obtained for  the nearest-neighbour version  of
the   Ising  model for a  range   of values  of  the  $q$ parameter of
Generalized Thermostatistics. We found  an evidence that the  2D-Ising
model does not undergo phase transitions at finite temperatures except
for the extensive case $q=1$.

\end{abstract}


\end{frontmatter}

\section{Introduction}

Many  systems  seem  to   be    well  described by a     non-extensive
thermostatistic   rather than   the  usual Boltzmann-Gibbs  statistics
(BGS).  When  the  effective microscopic interactions and  microscopic
memory    are   short-ranged    and  the    boundary    conditions are
non-(multi)fractal \cite{tsallis98b},  the BGS provides a complete and
consistent  description  of  the  system.   Otherwise, it   fails.  An
alternative  approach is the use  of  the so-called Tsallis statistics
(TS) \cite{tsallis88,tsallispage}.  The entropy  in the TS  is defined
as \cite{tsallis88}

\begin{equation}
S_q=k\frac{1-\sum_{i=1}^{\Omega}{p_i^q}}{q-1},\label{sq}
\end{equation}

with  $\sum_i{p_i} =1  $,   where $i$ is    a given state  with energy
$\epsilon_i$   from  $\Omega$ possible states  and   $k$ is a positive
constant.  The index  $q$ characterizes the degree of  nonextensivity.
The limiting case $q=1$ recovers the usual BGS entropy definition. The
other contraints  needed to obtain  the thermodynamical  averages have
been     recently discussed     by  Tsallis,    Mendes   and  Plastino
\cite{tsallis98}.

Magnetic systems \cite{silva96}-\cite{andricioaei97} are to be counted
among the large variety  of  systems to   which  TS has  already  been
applied   \cite{tsallispage}.  The  reason  for  such  an  interest is
obvious: statistical physicists  have spent a lot of  time in the past
developing tools  to understand the critical  phenomena  presented by
this kind  of  systems - and have succeeded in this task.  As  a
natural consequence,  one could expect   some attempts at generalizing
well established   approaches developed for   extensive  systems.  The
real space renormalization group  \cite{cannas96a} and  the mean field
approximation\cite{nobre95}  are   examples of tools   that  have been
considered  in these generalizations.  However,   it is not surprising
that results from these  approximations seem to be controversial: even
fundamental questions such as how to  define and to obtain the correct
expression for the  thermodynamical averages have been  revisited very
recently\cite{tsallis98}.

The Monte Carlo method  is a powerful  tool frequently used in the BGS
framework  in  order to solve ambiguities of this sort.  However, within
the
TS framework,  Monte  Carlo simulations of   magnetic
systems  have not  been    exploited  so  far because  of    technical
difficulties that  we are going to address  in this work. One issue is
how to   the define the  acceptance  probabilities which  lead  to the
correct   distribution  of probabilities.    A  first alternative  was
presented in  the application of  the  generalized simulated annealing
for the Traveling Salesman Problem\cite{penna95}, where the acceptance
probability is a simple generalization of the Metropolis algorithm for
$q\neq 1$.   This alternative was  motivated by  the definition of the
thermodynamical   averages  in use   at that   time,  which imposed an
undesirable dependence on  the definition  of  the zero of  the energy
scale.  The use of the  above mentioned alternative for the acceptance
probabilities   conveniently   removes  this   dependence.    A second
alternative  was     proposed   by   Andricioaei   and  Straub    (AS)
\cite{andricioaei97} a  couple of  years after the   first one.  A new
expression for the   acceptance  probabilities was obtained  from  the
detailed balance condition (a sufficient  but not necessary  condition
for thermodynamical   equilibrium).   This  new  alternative  cleverly
circumvents the ambiguity  on the  definition  of the  lowest level of
energy.

The second reason why, in our opinion, Monte Carlo simulations are not
frequently used within the TS is the fact that all the
computational effort  involved  in a  BGS simulation  has to be  spent
again for each value of the parameter  $q$.  Since $q$ is a continuous
variable,  computer  simulations  are much    more
time-consuming  in the TS than in the BGS, if one  wants  to investigate
the $q$  dependence  of the
results.  That  is why it has  been difficult, for instance, to answer
the question about  which acceptance probabilities  to use in  a Monte
Carlo simulation of magnetic systems.  One of the goals of the present
work is  to  show that  the  recently proposed Broad  Histogram Method
(BHM) \cite{pmco96}  is  the ideal tool   for Monte Carlo  simulations
within the TS framework.  Because the $q$-independent
density of  states $g(E)$ is, in the  BHM, directly obtained from some
measured  microscopic  quantities,    we  are  able   to  obtain   any
thermodynamic  observable for all values of  $q$ and $T$ from only one
computer   run.    This fact  turns  BHM   simulations on  the TS
, {\em for all values of $q$}, as fast as in the BGS.

Briefly, in this paper we are suggesting the BHM as the ideal tool for
computer  studies  on the TS.    We show it  through a
simulation of  the   two-dimensional  Ising  model  with   short-range
interactions. We are aware of the fact that the  model we are going to
study  is an extensive      one,  therefore well  described   by   the
BGS. Among the reasons  why we decided to study
this system are:

\begin{itemize}
\item this model can be very easily simulated with great efficiency;
\item the exact solution for  the density of states of finite
systems is known in the limit $q=1$ \cite{beale96}; 
moreover, the BHM is able to reproduce this exact solution with great
accuracy \cite{pmco96};
\item  previous results   for this  model  using other   approximation
methods  are controversial and/or  inconclusive; our simulations could
shed some light on this ongoing discussion;
\item we  could  easily   and reliably  show  which choice    for  the
acceptance   probabilities   reproduces  the  Tsallis distribution  of
statistical weights \cite{penna95,andricioaei97};
\item this simple system is here used as a testing ground for the 
methods we propose; building on this first step allows us to use the 
same approach to study , other more complex non-extensive systems such as
the Ising model with long-range interactions.
\end{itemize}

In summary, our choice of a well known system is most convenient 
since we   are dealing with  two  brand-new recipes:  the
TS with normalized q-expectation  values and the BHM.

This paper is organized as follows: in the next section, we review the
TS  with normalized  q-expectation   values.  We also
include a discussion about the stability of the solutions for the free
energy in  this  new formalism.   In section  3,  we  review the Broad
Histogram Monte  Carlo Method and present its implementation  for the
TS. This   is followed by a presentation   of our results and
conclusions.

\section{The Tsallis Statistics with ``normalized q-expectation values''}

Tsallis, Mendes and  Plastino \cite{tsallis98} have recently discussed
the role of constraints within TS.  In that work, they
study three different alternatives for the internal energy constraint.
The first two choices  correspond to the ones  which have been applied
to many different systems in the last  years \cite{tsallispage}.  They
are:  (ia)    $\sum_i{p_i  \epsilon_i}=U$    and  (ib)   $\sum_i{p_i^q
\epsilon_i}=U_q$.  However,  both constraints present difficulties.  A
third   choice for  the    internal energy  constraint is defined   as
\cite{tsallis98}

\begin{equation}
\label{newcons}
U_q=\frac{\sum_{i=1}^{\Omega}{p_{i}^{q}\epsilon_i}}
{\sum_{i=1}^{\Omega}{p_{i}^{q}}}, \label{constrain}
\end{equation}

where $q$   is  the degree   of  non-extensivity, also present   in the
definition of entropy    (eq.  (1)).  Each  constraint   (ia),(ib) and
eq.(\ref{constrain}) determines a different set of probabilities $p_i$
for each state with  energy  $\epsilon_i$.  The extremization  of  the
generalized  entropy (\ref{sq}),  under  constraint  (\ref{constrain})
gives us an implicit equation for the probabilities $p_i$:

\begin{equation}
p_i=\left[1-\frac{(1-q)\beta(\epsilon_i-U_q)}
{\sum_{j=1}^{\Omega}{p_j^{q}}}\right]^{\frac{1}{1-q}}/Z_q\label{piq}
\end{equation}

with

\begin{equation}
Z_q(\beta)                     \equiv              \sum_{i=1}^{\Omega}
\left[1-\frac{(1-q)\beta(\epsilon_i-U_q)}
{\sum_{j=1}^{\Omega}{(p_j)^{q}}}\right]^{\frac{1}{1-q}}
\end{equation}

The  normalized q-expectation  value  of  an observable is   therefore
defined as

\begin{equation}
\label{expectation}
O_q        \equiv        \langle       O_i        \rangle_q     \equiv
\frac{\sum_{i=1}^{\Omega}{p_i^q O_i}}{\sum_{i=1}^{\Omega}{p_i^q}}
\end{equation}

where $O$  is  any observable which commutes  with   the Hamiltonian -
otherwise we should make  use of the  density operator $\rho$.  We
will refer  to this reformulation of  the TS as ``with
normalized q-expectation  values''.  A  very important consequence  of
this new  definition of constraints  is that the probabilities  do not
depend on the choice of the zero of energy.

In order  to solve eq.  (\ref{piq}) Tsallis  {\it et al.}  suggest two
different approaches,  namely the {\em   Iterative Procedure} and  the
{\em  $\beta \rightarrow  \beta'$  transformation}.  In  the iterative
procedure,  we start with an  initial set of probabilities and iterate
them  self-consistently until the desirable  precision is reached.  In
the  $\beta  \rightarrow \beta'$ transformation  the  set of equations
above is transformed to:

\begin{equation}
p_{i}  = \left[1-(1-q)\beta'\epsilon_{i}\right]^{\frac{1}{1-q}}/Z'_{q}
\\
\end{equation}
 
\begin{equation}
Z'_{q}\equiv \sum_{j=1}^{\Omega}{\left[1-(1-q)\beta'\epsilon_j\right]^
{\frac{1}{1-q}}}\\
\end{equation}

with

\begin{equation}
\label{betabetaast}
\beta'(\beta)\equiv              \frac{\beta}{             {(1-q)\beta
U_{q}+\sum_{j=1}^{\Omega} p_j^q}}.
\end{equation}

This set  of equations is  similar to  the one  that is obtained using
constraint (ib), except    for  its dependency on   the   renormalized
temperature, given by eq. (\ref{betabetaast}).

In order to obtain ${p_i}$, we go through the following steps:

\begin{enumerate}
\item   Compute the quantities $y_i=(1-(1-q)\beta'\epsilon_{i}),  \;\;
\forall i\in\Omega $;
\item If $y_i<0$ them $y_i=0$;
\item Compute $Z_{q'}=\sum_{i=1}^{\Omega}{y_i^{1/(1-q)}}$;
\item Compute $p_i(\beta')=y_i^{1/(1-q)}/Z_{q'}$;
\item Obtain  $U_q(\beta')$  and  any other  thermodynamical  quantity
using eq. (5);
\item Obtain $\beta(\beta')$ from equation (\ref{betabetaast}).
\end{enumerate}

This recipe  allows    the determination  of   $p_i(\beta)$  for   all
$\beta(\beta')$     and consequently  $U_q(\beta)$    (and  any  other
observable). The second step in the above  procedure is the well known
cutoff     \cite{tsallis98}    associated    to     {\em   ``vanishing
probabilities''}.   This cutoff is required   only for $q<1$.  Because
the  cutoff is  applied    before  the   actual computation   of   the
probabilities, the norm constraint is still respected.

It has   been  shown \cite{lima98} that   both  recipes, the iterative
method and  $\beta\rightarrow\beta'$ transformation,  demand a careful
analysis  before its application.  The  free  energy obtained from the
iterative procedure presents a  non-physical discontinuity whereas the
free energy from  the  $\beta \rightarrow \beta'$   transformation has
loops. To get rid of  these pathologies, one  has to make explicit use
of the  minimization   condition on  the   free  energy, whenever   an
ambiguity appears.     This procedure generates  the  correct internal
energy and temperature dependency and restores  the proper behaviour of
the thermodynamic    observables.     Following the   suggestion    of
ref.\cite{lima98}, we use in this paper the $\beta \rightarrow \beta'$
transformation with the proper corrections.

\section {The Broad Histogram Monte Carlo Method}

The approach that we  are going to discuss in  this section, the Broad
Histogram Method (BHM) \cite{pmco96}-\cite{pmco98}, is one of the many
attempts  at   doing  very  efficient  simulations.   In   traditional
simulations,  we  need a new run  for  each  value of the temperature.
However, some  different  approaches have  been proposed in  which  we
compute some quantities at a given temperature and reweight them for a
different  temperature  (see, for instance, ref.~\cite{marinari96} and
references therein). One of these approaches  is the histogram method,
first  introduced  by  Salzburg \cite{salzburg59}  and  popularized by
Ferrenberg  and Swendsen \cite{ferrenberg88}.    However, it  has been
shown  \cite{ferrenberg91}  that  the  histogram method  presents some
limitations, the most  important concerning the range of  temperatures
for which just    one run is  sufficient.    The BHM enables us to
directly calculate  the energy spectrum  $g(E)$, without any need for a
particular    choice  of   the   thermostatistics     to  be   used
\cite{pmco96}-\cite{pmco98}.

In the BHM the energy degeneracy is calculated through the steps:

{\em Step  1}:  Choice  of a  micro   reversible protocol of   allowed
movements in the state space of the  system such that changing from an
$X_{\rm old}$  to an $X_{\rm  new}$ configuration  is  allowed if, and
only  if, the reverse  change is  also  allowed (the protocol  must be
micro-reversible):

\begin{equation}
\underbrace{X_{\rm   old}  \rightarrow   X_{\rm  new}}_{\rm   allowed}
\Longleftrightarrow   \underbrace{X_{\rm   new}   \rightarrow   X_{\rm
old}}_{\rm allowed};
\end{equation}

it is important  to note that these movements  are virtual, since they
are not actually performed.

{\em Step 2}: Choice of a fixed amount of energy change $\Delta E_{\rm
fix}$  and computation of  $N_{\rm  up}(X)$ ($N_{\rm dn}(X)$) for  the
configuration X, defined as the number of allowed movements that would
increase (decrease) the energy of  the configuration by $\Delta E_{\rm
fix}$.    Then   $\langle    N_{\rm   up}(E)\rangle$  ($\langle N_{\rm
dn}(E)\rangle$) is the  micro  canonical  average of $N_{\rm   up}(X)$
($N_{\rm dn}(X)$) at energy $E$;

{\em Step 3}: Since the total number  of possible movements from level
$E+\Delta  E_{\rm fix}$ to  level $E$ is  equal to the total number of
possible movements from level $E$  to level $E+\Delta E_{\rm fix}$, we
can write down the equation

\begin{equation}
\label{bhrel}
g(E)\langle N_{\rm  up}(E)\rangle   =  g(E+\Delta E_{\rm  fix})\langle
N_{\rm dn}(E+\Delta E_{\rm fix})\rangle .
\end{equation}

This  relation is  exact  for  any statistical  model  or  any  energy
spectrum \cite{pmco98}. It can be rewritten as

\begin{equation}
\ln  g(E+\Delta E_{\rm fix}) -   \ln g(E) =  \ln \frac{\langle  N_{\rm
up}(E)\rangle}{\langle N_{\rm dn}(E+\Delta E_{\rm fix})\rangle}
\end{equation}

If we choose  $\Delta E_{\rm  fix} \ll  E$, the above  equation can be
approximated by

\begin{equation}
\frac{\d   \ln  g(E)}{\d     E}=\frac{1}{\Delta   E_{\rm  fix}}    \ln
\frac{\langle N_{\rm up}(E)\rangle}{\langle N_{\rm dn}(E)\rangle}
\end{equation}

This equation can be easily solved for  $g(E)$.  Once this quantity is
known, the expected value of some observable $O$ can be calculated by

\begin{equation}
\label{qexpectation}
\langle  O  \rangle _{q,T}   =  \frac{\sum_E{\langle O \rangle_E  g(E)
[p(E)]^q}}{\sum_E g(E) [p(E)]^q}
\end{equation}

This method   (in the q=1   BGS) was  first applied to   systems  with
discrete   degrees of freedom.   Recently  it   has been extended to
continuous  systems,  such as the   XY Model  \cite{munoz98}.   To our
knowledge, this is the first time the method is being applied to a
different
statistics. Besides the more accurate and faster results in comparison
to traditional  methods, the  BHM is  even  more  efficient in  this
particular  case  because  eq.  (13)  is  the   only  quantity  to  be
recalculated for each new value of $q$.

\section
{Implementation of the BHM on the 2D Ising Model with first
neighbour interactions}

To clarify  the application of our ideas  to magnetic systems, we will
use  the square lattice ferromagnetic Ising Model with first
neighbour interaction. The Hamiltonian for this system is given by

\begin{equation}
{\cal H} \equiv H/J = -\sum_{\langle ij \rangle}{\sigma_i\sigma_j}
\end{equation}

where  $\sigma_i= \pm  1$, $J$  is a  positive  constant.  The  sum is
performed over all  pairs of first neighbours  in a  square lattice of
size $N  = L \times L$.   For an efficient  implementation, we rewrite
the Hamiltonian as

\begin{equation}
{\cal H}_m  \equiv H_m/J =  \sum_{\langle ij  \rangle}{\zeta_i \otimes
\zeta_j} = \frac{-\sum_{\langle ij \rangle}{\sigma_i\sigma_j}+2N}{2}
\end{equation}

where $\zeta_i= 0$  or $1$ and  $\otimes$ represents  the exclusive OR
operation, where $0 \otimes 0 = 1 \otimes  1 = 0$ and  $1 \otimes 0 = 0
\otimes 1  =  1$.  The    critical  temperature  for $q=1$   in    the
thermodynamical  limit for  this renormalized   Hamiltonian is $T_c  =
{[2{\rm arctanh}(\sqrt{2}-1)]}^{-1}=1.13459...$.

We choose a single spin flip  protocol  of movements to obtain
$\langle N_{\rm up}(E)\rangle$ and  $\langle N_{\rm dn}(E)\rangle$. As
proposed in  \cite{pmco96} and \cite{pmco97},  an unbiased random walk
is them performed on the energy axis in order  to visit all values on
the energy spectrum.

The   number  $N_{\rm up}(E)$ ($N_{\rm     dn}(E)$) of movements  that
increase (decrease) the current value of energy  $E$ by $\Delta E_{\rm
fix}$  is used to calculate  $g(E)$.  The magnetization $M(E)$ is also
stored for  each lattice size.  Here  we choose $\Delta E_{\rm fix}=4$
out of the possible  values $|\Delta E|$=$0,  2, 4$.  The histogram is
obtained from 6.400.000 samples  (distributed in the energy  axis) for
L=30,50,70,100.  Obtaining the histograms for  $L=30$ takes 20 minutes
of CPU time on a DEC Alpha 400. For $L=100$, the time increases to 220
minutes. 

The next step is to obtain the set of probabilities using the $\beta
\rightarrow  \beta'$  procedure.  After this    step, we can use  eq.
(\ref{qexpectation})  to obtain the  internal  energy $U_q(T)$ and the
magnetization $M_q(T)$.  The  free energy $F_q(T)$ is  also calculated
using

\begin{equation}
F_q \equiv U_q - T S_q = U_q - \frac{1}{\beta}\frac{Z_q^{1-q}-1}{1-q}
\end{equation}

Instead of using eq.   (4), it is  more efficient to use the  relation
\cite{tsallis98}

$$ Z_q^{1-q}=\sum_{i=1}^\Omega{p_i^q}=\sum_E{g(E)[p(E)]^q}
$$

The CPU time for the determination of the  values of any observable in
the   whole  range  of  temperatures is    independent  of the lattice
size. Typically, it takes  30s on a DEC Alpha  400 for 20000 values of
temperature, for each value of q.

For    the   sake of  comparison,     we  have  implemented  multispin
\cite{pmcobook}  versions of the  Ising  Model with the  AS acceptance
probability \cite{andricioaei97}

\begin{equation}
p=\frac{1}{2}[1-\tanh(\beta'\Delta\bar{E}/2)]
\end{equation}

where $\bar{E}=[1/\beta'(q-1)]\ln [1-(1-q)\beta' E]$, and with the TSP
acceptance probability \cite{penna95}

\begin{equation}
p=(1-(1-q)\beta'\Delta E)^\frac{1}{1-q}.
\end{equation}

Notice that the equations above  are written as functions of  $\beta'$
instead of $\beta$. The reason  is that they  were proposed before the
publication  of  the solution for the  constraints  problem in  the TS
\cite{tsallis98} discussed in section 2 .

For an $L=30$ lattice, a simulation  run, using the traditional method
with  the   same numer   of Monte Carlo    steps as  used  in  the BHM
simulation, took 60s of  CPU  time {\bf for  a  single pair of  values
$(q,T)$}. This should be compared to the 1200s of BHM  for all values of
q and the whole range of temperatures.

\section{Results}

We will now discuss the results obtained from the
implementation described  in  the previous section.   We begin  with a
comparison between the  results obtained  through  the use of  the two
already mentioned techniques:   the BHM and  the traditional multispin
simulation with both the AS and the TSP probabilities.  The comparison
must be made  in terms of  $T'$, because the  AS and TSP probabilities
were derived  before   the publication of Ref.    \cite{tsallis98}, as
pointed above. In Fig. 1 we show the results for the magnetization and
the internal energy as functions of $T'$. 

\begin{figure}[!h]
\epsfysize=14cm \rotate[r]{\epsfbox{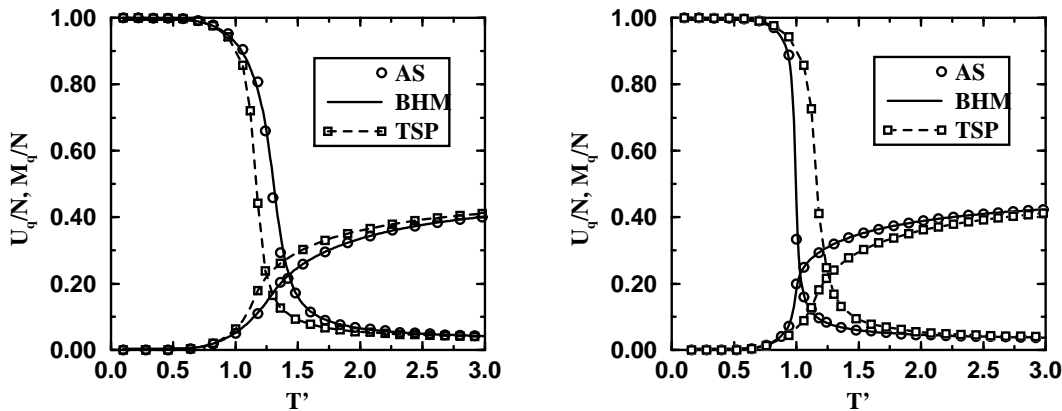}}
\caption{Comparison  between the  results  from  the  BHM, AS and  TSP
approaches. In both the AS and TSP simulations, 1000 samples were used
for  each $T$.  The BHM  and AS probabilities  agree in both $q<1$ and
$q>1$  regime.  The TSP probabilities deviate  from these results even
for  values of  $q$  very close  to one  (where  the three results are
expected to agree).} 
\end{figure}

The remarkable agreement between the  results coming from both the BHM
and    the AS  techniques contrast   with  those coming   from the TSP
probabilities.  Therefore, this last one should be discarded as an
approach to
the simulation    of physical systems    in   the TS 
framework.  In Fig. 1 the values for  $q$ were chosen  to be close to
one.

We are going to discuss separately the $q<1$ and $q>1$ regimes, since
the system behaves differently in these regions.

\subsection{The $q<1$ regime}

Fig. 2 shows the internal energy as a function of $T$, for some values
of  $q$ very close  to  one.  The first  surprising result  is that a
reentrant region develops as soon as $q$ gets smaller than $1$.
This
reentrant behaviour  is even    more  noticeable as the   lattice  size
increases,  for the  same  value of  $q$.  This   behaviour is already
being reported \cite{lima98} for a two-level system. We are going to
use  the same recipe proposed therein to deal with this
pathology. 

The reentrant behaviour is also present in the free energy curve, as we
can see  in Fig.  3.   To reconstruct the  correct  curve for the free
energy  and, as a consequence, for   any   thermodynamical   macroscopic
quantity,  we have to choose a  criterion to discard  two of the
three possible  states  in the  loop (see  the inset of   Fig.3).  We
choose to consider only the state that corresponds to the lowest value
of the free energy. The correct curve for  the internal energy is also
displayed in   Fig.3, and is another illustration that supports  
the statement that   TS  with renormalized q-expectation
values is thermodinamically    stable.   A deeper discussion   of  the
reentrant behaviour and the  technique developed to  restore uniqueness
is being published elsewhere \cite{lima98}. 

\begin{figure}[!h]
\epsfysize=14cm \centerline{\rotate[r]{\epsfbox{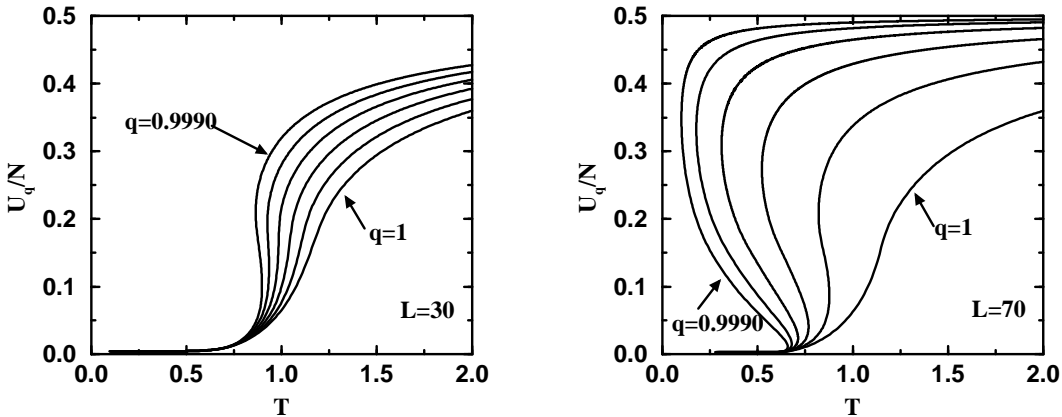}}} 
\caption{Internal energy for $L=30$ and $L=70$.  The curves (from left
  to  right in all graphs)   correspond to  different $q$ values ranging 
from 0.9990 to  1. The reentrant behaviour is being reported to appear
  in a two-level system.} 
\end{figure}

\begin{figure}[!h]
\epsfxsize=7cm \centerline{\rotate[r]{\epsfbox{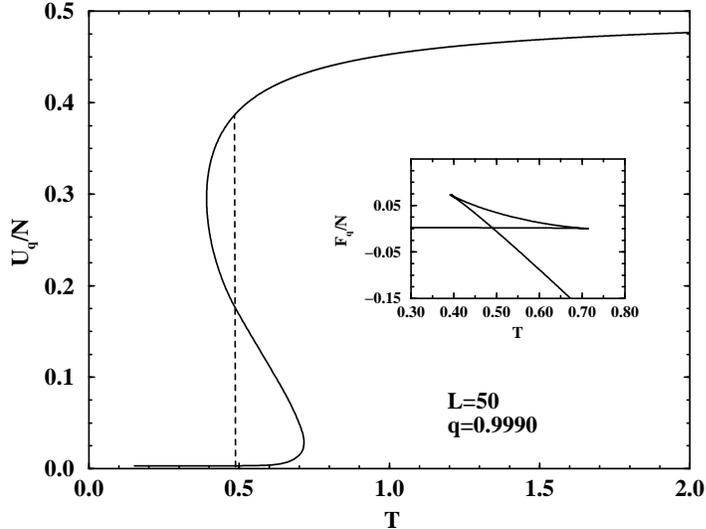}}}
\caption{Internal energy as a function of the temperature for $L=50$ and
 $q=0.9990$.  The dashed  line  shows   the correct behaviour     when
non-uniqueness  is removed by stability  arguments.  The
inset shows the  loop  in the  free  energy  that corresponds to   the
reentrant region,  the  removal  of  which  restores  uniqueness;  the
resulting internal energy is left with a discontinuous derivative with
respect to temperature.  }
\end{figure}

When uniqueness is  restored by removing  the  loop, there results a 
free energy with a discontinuous  first derivative  with respect to
temperature  at  a point which  could  be identified as the transition
temperature.  This means that the entropy has a discontinuity at
this temperature. In addition, the magnetization, as  shown in Fig. 4,
also presents a discontinuity  at this point  - after the reentrancies
have been removed.  These discontinuities become more noticeable as
we consider ever smaller values of $q$ (see Fig. 4) and/or bigger
lattices  (see Fig.  5). All these results were obtained for finite
lattices. To determine the transition temperature in the thermodynamic
limit  $L \rightarrow \infty$, we plot  the transition temperature for
finite lattices as a function of its inverse size, $1/L$, and take the
limit $1/L \rightarrow 0$. 

\begin{figure}[!h]
\epsfysize=14cm \centerline{\rotate[r]{\epsfbox{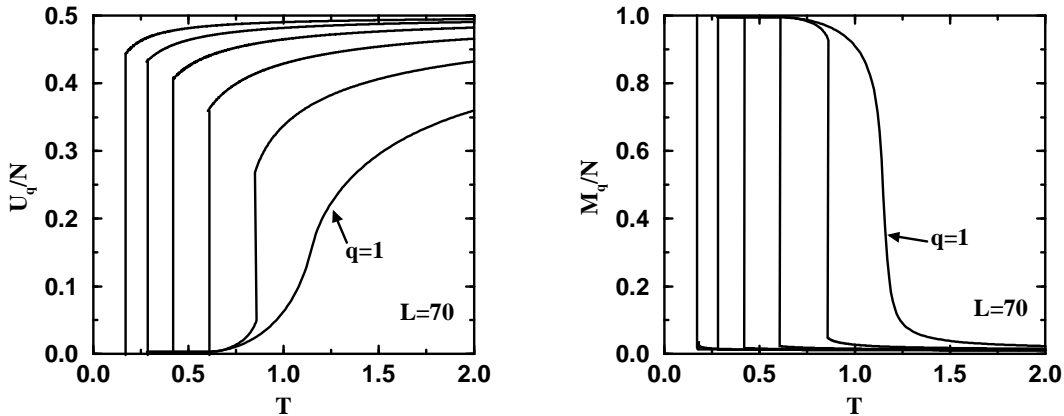}}} 
\caption{Curves  obtained after the  correction procedure discussed in
the  text. The plot of the internal energy on the left correspond to
the right part of
Fig. 2. On the right, we show the magnetization. The values of $q$ are
the same as in Fig. 2.} 
\end{figure}

\begin{figure}[!h]
\epsfysize=14cm \centerline{\rotate[r]{\epsfbox{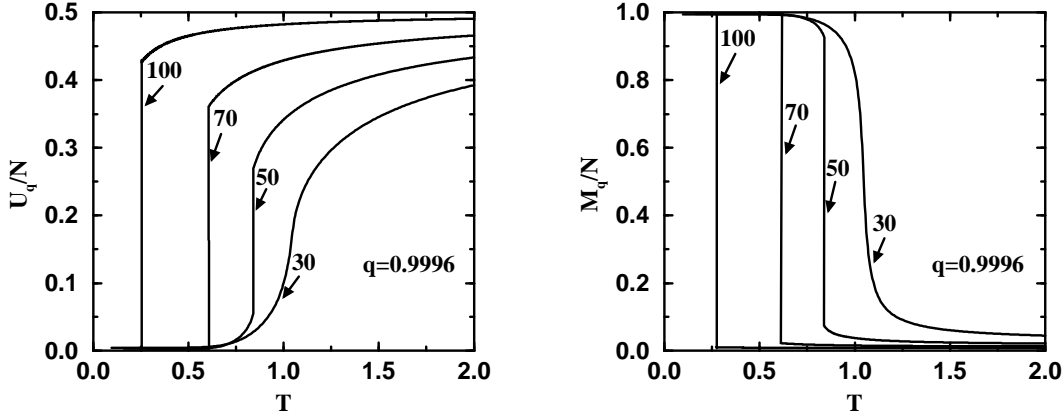}}} 
\caption{Internal energy and magnetization as functions of temperature,
now for a fixed value of $q$ and different
lattice sizes.} 
\end{figure}

\begin{figure}[!h]
\epsfxsize=7cm \centerline{\rotate[r]{\epsfbox{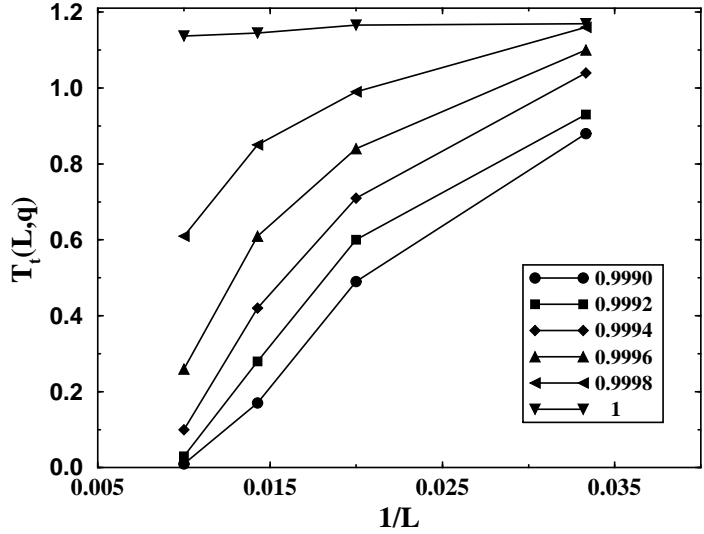}}} 
\caption{Determination of the  transition temperature.  The  lines are
only guides to the eyes.} 
\end{figure}

Fig. 6 shows an attempt of  finite-size scaling for different values
of $q$.  For $q<1$  and $L\rightarrow\infty$, the critical temperature
vanishes extremely fast. Our results point out  that there is no phase
transition for $q<1$ in the thermodynamic limit.

\subsection{The $q>1$ regime}

Basically, the same approach we have used for $q<1$ is also used here. 
However, we have found that the 2D Ising model presents a quite different
behaviour in both regimes (actually, it is also different from the $q=1$
case). Fig.  7 shows the internal energy and magnetization as functions of
$T$ for some values of $q$ on a $L=70$ lattice. From the internal energy
curve, we promptly find that the discontinuity in its derivative (specific
heat) at $T=T_c$ is not present anymore, therefore there are no evidence
of a phase transition. This is also clear from the behaviour of the
magnetization as a function of temperature.  Fig.  8 supports this
conclusion; the magnetization gets smoother with increasing values of $q$
or the lattice size.

\begin{figure}[!h]
\epsfysize=14cm \centerline{\rotate[r]{\epsfbox{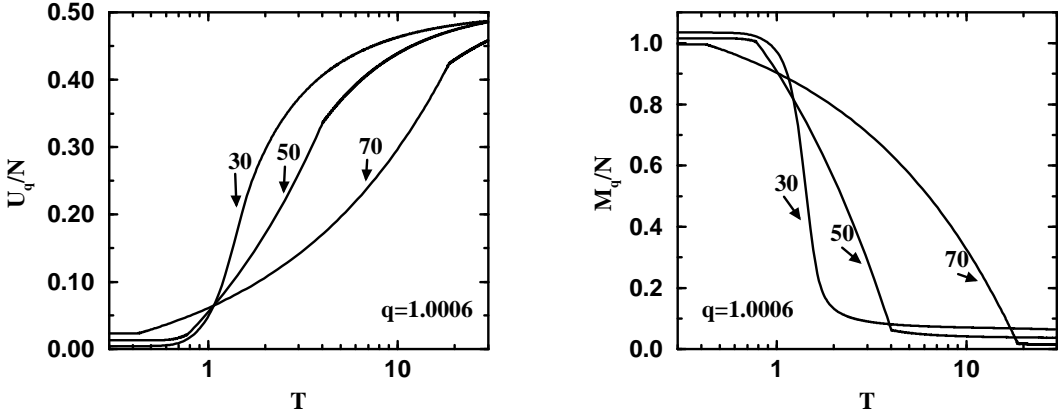}}}
\caption{Internal energy and magnetization as functions of temperature
for $L=70$ and $q=1$ to 1.0006. The boxes are used to display regions 
where abrupt changes in the derivative of the functions occur. The curves
on this and the next figures are vertically displaced to allow better
examination of these regions.} \end{figure}

\begin{figure}[!h]
\epsfysize=14cm \centerline{\rotate[r]{\epsfbox{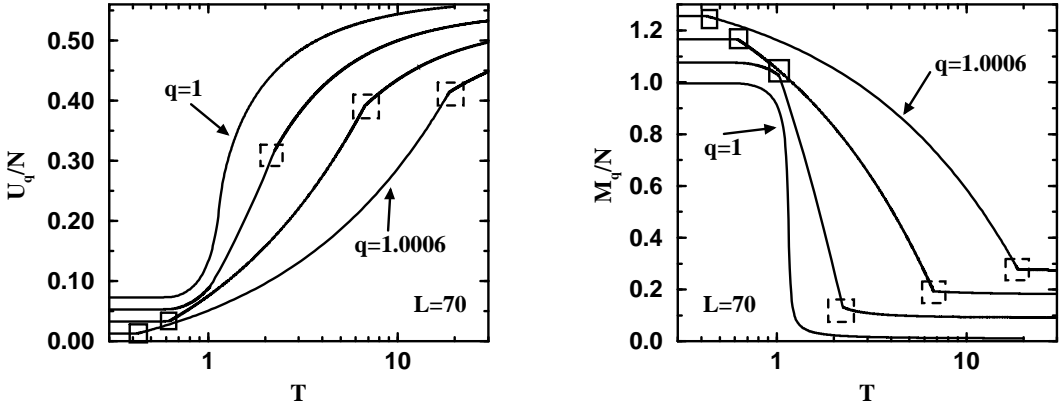}}}
\caption{Internal energy and magnetization as functions of temperature
for a fixed value of $q=1.0006$ and for different lattice sizes. Again, 
the boxes are used to display regions where abrupt changes in the 
derivative of the functions occur.}
\end{figure}

A careful examination of Fig. 7 and Fig. 8 reveals discontinuous changes
in the derivative of the curves at  two points (these changes are more
easily   seen for  the   largest  values   of $q$   and   $L$).  These
discontinuous  derivatives  are  related   to the  transformation   $T
\rightarrow T'$. Fig. 9 presents the relation between $T$ and $T'$ for
the same values of $L$ used in Fig. 8 and for a range of values of $q$
slightly larger than what  was used in  Fig.  7.  The curves display a
region of abrupt change (but not a discontinuity) which gets larger as
$q$ is increased,  for a fixed  $L$. The values  of T that limit  this
region are also those that lead  to a discontinuous specific heat, and
appear to  have zero and $\infty$ as  limits when $q$ or $L$ increase.
In  the thermodynamic  limit,  the  magnetization is  always  positive
(therefore,  the system is always in  the ordered state, except at the
limit of infinite temperature) and  the derivative of the specific
heat is always positive.  Therefore,  we argue that  there is no phase
transition for the 2D Ising model, for $q>1$.

\begin{figure}[!h]
\epsfysize=14cm \centerline{\rotate[r]{\epsfbox{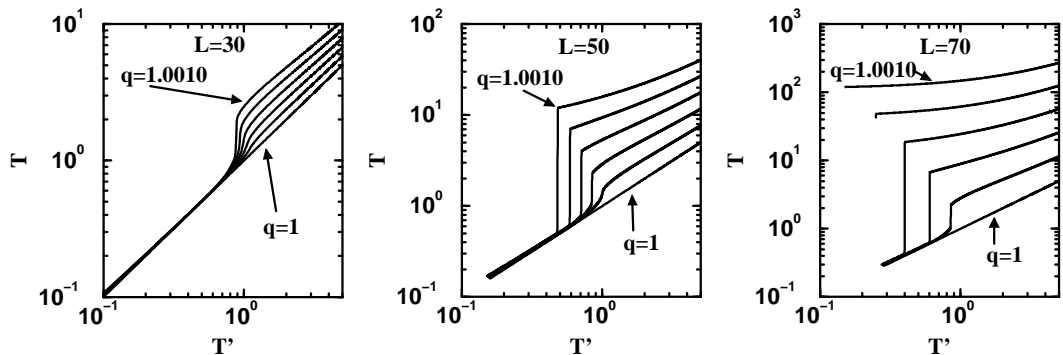}}}
\caption{$T$ as a function of $T'$, for different lattice sizes.   The
curves correspond to different values of $q$ ranging from 1 to 1.001.}
\end{figure}

\section{Conclusion}

We studied  the   simulation  of  magnetic  systems   in  the  Tsallis
Statistics  using the Broad Histogram  Method.  It was shown that this
method is very  efficient,   since all thermodynamic   observables of
interest  can be calculated  for  a  new value   of the  parameter $q$
without  the need for a  new computer  run.   The square lattice Ising
model with nearest neighbour interactions was chosen as an example to test
the method.  All previous work on the nearest neighbour Ising model
suggest that there is a phase transition for each value of $q \neq 1$.
However, our results show that there is no transition in this model
for any value of $q \neq 1$. Only the $q=1$  system presents the usual
second order phase transition.

Further step  along these lines would be the simulation of spin models 
with long-range interactions.  We believe that
these intrinsically non-extensive systems
can  display   a  richer   behaviour,  within  the  Tsallis  Statistics
framework, than  the model studied in  this  work.  Actually, previous
studies   of magnetic systems  in  the Tsallis Statistics have already
suggest that this is the case. However, a powerful simulational tool 
such as the Broad Histogram Monte Carlo Method used in this work was
lacking and only recently has become available. The demonstration that we 
now present of its usefulness will surely bring considerable advance  
in  the understanding of the behaviour of magnetic systems in a
non-extensive regime.

\section*{Acknowledgments}

The authors are indebted to Professor C.  Tsallis for extensive ($q=1$!) and
enlightening discussions. This work was partially supported by CNPq and CAPES
(Brazilian Agencies).

\end{document}